\documentclass[acmtog,table,nonacm]{acmart}

\usepackage[ruled,linesnumbered]{algorithm2e} 

\usepackage{subfigure}
\usepackage{epstopdf}
\usepackage{wrapfig}
\usepackage{multirow}
\usepackage{mathrsfs}
\usepackage{colortbl}
\usepackage{BOONDOX-cal}
\usepackage{url}
\usepackage{tikz}
\usepackage{makecell}
\usepackage{hyperref}
\usepackage{cleveref}
\usepackage{todonotes}
\usepackage{mathtools}
\usepackage{listings,color}
\usepackage{xcolor} 
\definecolor{listinggray}{gray}{0.9}
\definecolor{lbcolor}{rgb}{1,1,1}
\definecolor{Darkgreen}{rgb}{0,0.4,0}
\Crefname{equation}{Eq.}{Eqs.}
\Crefname{figure}{Fig.}{Figs.}

\AppendGraphicsExtensions{.tga}

\begin{document}
\setcopyright{none} 
\author{Tianyu Wang}
\affiliation{\institution{Independent} \country{China}}
\email{wtyatzoo@zju.edu.cn}

\title{Differentiable 3D Triangle-Triangle Intersection Energy}

\begin{abstract}
        Obtaining intersection-freeness or global injectivity is important in computer graphics.
However, it is challenging, especially for the non-oriented deformation primitives.
Most methods often rely on an intersection-free initialization and track the continuous trajectory to keep the legitimacy and cannot be used for the task without such an initialization.
For the latter one in 3D space, we introduce a novel second-order differentiable energy defined from the 3D triangle-triangle intersection testing,
and a GPU-based inexact Newton optimization route.
We show that intersection can be efficiently resolved 
integrated with our method,
requiring no user interaction, history information or a valid initialization.
\end{abstract}


\maketitle

\section{3D Triangle-Triangle Intersection Energy}
\label{sec:ttie_def}
Our method is built on the M{\"o}ller's method~\cite{moller1997fast}. The key observation is that
we can reduce the 3D triangle-triangle intersection test to the overlapping test of two 1D line segments.
Therefore, it is possible to define a simple closed-form differentiable energy to resolve this overlapping. 
Once the overlapping is resolved, the triangle-triangle intersection should also be resolved.

\begin{figure}[t]  
        \centering                
        \subfigure{\includegraphics[width=1.6in]{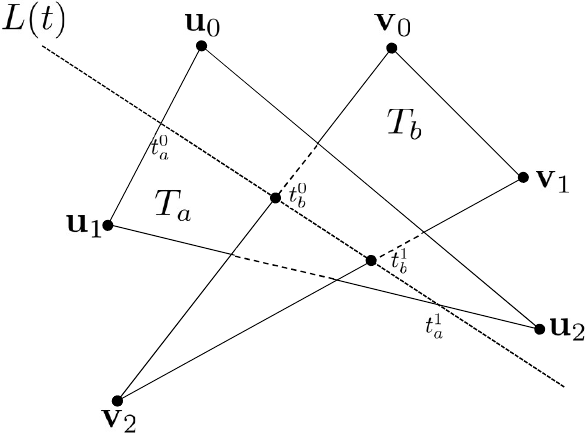}}
        \vspace{-0.06in}
        \caption{\textbf{M{\"o}ller's method for triangle-triangle intersection test in 3D.} }
        \vspace{-0.2in}        
        \label{fig:moller_method}        
\end{figure}

We consider the general situation where two intersected triangles $T_a$ and $T_b$ are not coplanar or topologically adjacent as shown in \Cref{fig:moller_method}:
$\textbf{u}_{0,1,2}$ are the three vertices of $T_a$ and $\textbf{v}_{0,1,2}$ are the three vertices of $T_b$; $T_a$ intersects with the plane of $T_b$ and forms a line segment
whose parameter interval $I_a$ on the straight line $L(t)$ is $[t_a^0,t_a^1]$; $T_b$ intersects with the plane of $T_a$ and forms a line segment
whose parameter interval $I_b$ on the straight line $L(t)$ is $[t_b^0,t_b^1]$. M{\"o}ller showed that $T_a$ intersects with $T_b$ if and only if $I_a \cap I_b \neq \emptyset$.
This can be further equivalent to that:
\begin{equation}
        0 \in I_c \coloneqq [t_c^0,t_c^1] = I_a \oplus (-I_b) = [t_a^0-t_b^1,t_a^1-t_b^0],
\end{equation}
as illustrated in~\cite{schneider2013convex,Minkowski2024}, where $I_a \oplus (-I_b)$ means the \emph{Minkowski sum} of the point set $I_a$ and the negated point set of $I_b$.
To ignore the critical situation first, we can say that $T_a$ intersects with $T_b$ if and only if $0 \in (t_c^0,t_c^1)$, i.e., $(0-\frac{t_c^0+t_c^1}{2})^2 \textless (\frac{t_c^1-t_c^0}{2})^2$.
Therefore, we define our triangle-triangle intersection energy $\mathcal{T}(\textbf{u},\textbf{v})$ as below:
\begin{equation}
   \mathcal{T}(\textbf{u},\textbf{v}) = ((0-\frac{t_c^0+t_c^1}{2})^2 - (\frac{t_c^1-t_c^0}{2})^2)^2 = (t_c^0(\textbf{u},\textbf{v}) t_c^1(\textbf{u},\textbf{v}))^2.
   \label{eq:ttie}
\end{equation}







\subsection{Gradient and Hessian}   
\label{sec:grad_hess}
To compute the gradient of $\mathcal{T}(\textbf{u},\textbf{v})$ w.r.t. $\mathbf{u},\mathbf{v}$ is straight forward:
\begin{equation}
        \frac{\partial \mathcal{T}(\textbf{u},\textbf{v})}{\partial (\textbf{u},\textbf{v})} = 2t_c^0 (t_c^1)^2 \frac{\partial t_c^0(\textbf{u},\textbf{v})}{\partial (\textbf{u},\textbf{v})} + 2t_c^1 (t_c^0)^2 \frac{\partial t_c^1(\textbf{u},\textbf{v})}{\partial (\textbf{u},\textbf{v})}.
        \label{eq:ttie_grad}
\end{equation}
Ignoring the asymmetric part of the exact Hessian, its Hessian matrix can be expressed as below:
\begin{equation}
\left\{        
\begin{array}{l}
        \frac{\partial^2 \mathcal{T}(\textbf{u},\textbf{v})}{\partial (\textbf{u},\textbf{v})^2} \approx \\
        2(t_c^1)^2 \frac{\partial t_c^0(\textbf{u},\textbf{v})}{\partial (\textbf{u},\textbf{v})} \otimes \frac{\partial t_c^0(\textbf{u},\textbf{v})}{\partial (\textbf{u},\textbf{v})} + \\
        2(t_c^0)^2 \frac{\partial t_c^1(\textbf{u},\textbf{v})}{\partial (\textbf{u},\textbf{v})} \otimes \frac{\partial t_c^1(\textbf{u},\textbf{v})}{\partial (\textbf{u},\textbf{v})} + \\
        4t_c^0t_c^1 (
                \frac{\partial t_c^1(\textbf{u},\textbf{v})}{\partial (\textbf{u},\textbf{v})} \otimes \frac{\partial t_c^0(\textbf{u},\textbf{v})}{\partial (\textbf{u},\textbf{v})} +         
                \frac{\partial t_c^0(\textbf{u},\textbf{v})}{\partial (\textbf{u},\textbf{v})} \otimes \frac{\partial t_c^1(\textbf{u},\textbf{v})}{\partial (\textbf{u},\textbf{v})} 
        ),   
\end{array}
\right.             
        \label{eq:ttie_hess}
\end{equation}
where $\otimes$ is the outer product symbol.
\paragraph{Eigenanalysis} \hspace{0.12in}\
\label{sec:eigen_analysis}
We further analyze the eigensystem of Eq.~\ref{eq:ttie_hess}. The inexact Hessian is composed of three parts: 
the first part has a non-zero eigenvalue, which equals $2(t_c^1)^2 \big\|\frac{\partial t_c^0(\textbf{u},\textbf{v})}{\partial (\textbf{u},\textbf{v})}\big\|^2$ and is always positive;
the second part has a non-zero eigenvalue, which equals $2(t_c^0)^2 \big\|\frac{\partial t_c^1(\textbf{u},\textbf{v})}{\partial (\textbf{u},\textbf{v})}\big\|^2$ and is always positive;
the third part has a non-zero eigenvalue, which equals $8t_c^0t_c^1 \frac{\partial t_c^0(\textbf{u},\textbf{v})}{\partial (\textbf{u},\textbf{v})} \cdot
\frac{\partial t_c^1(\textbf{u},\textbf{v})}{\partial (\textbf{u},\textbf{v})}$ and might be negative.

\subsection{GPU Implementation} \hspace{0.12in}\
\label{sec:gpu_imp}
The gradient and Hessian computation of our energy is built on an open-sourced implementation~\cite{schneider2002geometric,geometrictools2024} of M{\"o}ller's method.
Considering the performance and memory consumption, we do not use forward-mode or reverse-mode automatic differentiation techniques, but manually
construct the computation graph to reduce the number of multiplications, and then  record the related gradient to construct the final Jacobian matrix on GPU. 
A C++ version of the gradient and Hessian computation based on the autodiff library~\cite{autodiff} can be found in Appendix~\ref{sec:ttie_autodiff} for clarity. 
To accelerate the triangle-triangle intersection testing, we use spatial hashing structure~\cite{Pabst2010} to eliminate redundant computations.


\begin{figure}[t]  
  \centering                
  \subfigure{\includegraphics[width=3.2in]{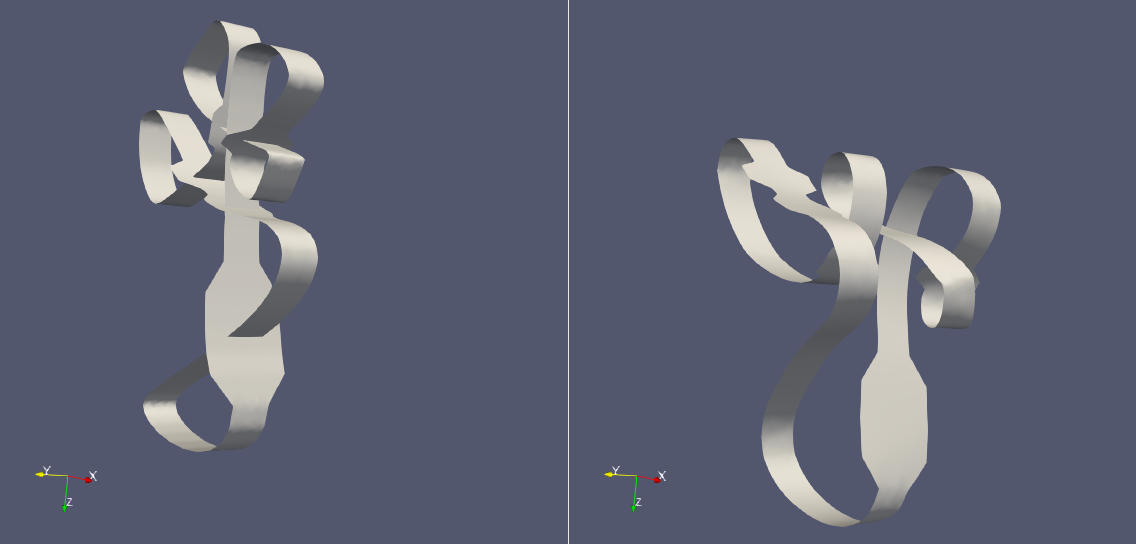}}
  \vspace{-0.06in}
  \caption{\textbf{A quasi-static simulation with our 3D triangle-triangle intersection energy, which evolves from an invalid state (left) to a valid one (right).} }
  \vspace{-0.2in}        
  \label{fig:res}        
\end{figure}








\bibliographystyle{ACM-Reference-Format}

\appendix
\section{C++ Code for our 3DTTI energy}
\label{sec:ttie_autodiff}
\begin{lstlisting}
bool intersects(
        Vector3dual& U0, Vector3dual& U1, Vector3dual& U2,
        Vector3dual& V0, Vector3dual& V1, Vector3dual& V2,
        Vector3dual& segment0, Vector3dual& segment1) {
  Vector3dual edge1 = U1 - U0;
  Vector3dual edge2 = U2 - U0;
  Vector3dual normal = edge1.cross(edge2);
  normal.normalize();
  dual d[3];
  int positive = 0, negative = 0, zero = 0;
  d[0] = normal.dot(V0 - U0);
  d[1] = normal.dot(V1 - U1);
  d[2] = normal.dot(V2 - U2);
  for (int i = 0; i < 3; ++i) {
    if (d[i].val > 0.0) {
      ++positive;
    } else if (d[i].val < 0.0) {
      ++negative;
    } else {
      ++zero;
    }
  }
  // positive + negative + zero == 3
  if (positive > 0 && negative > 0) {
    if (positive == 2)  // and negative == 1
    {
      if (d[0].val < 0.0) {
        segment0 = (d[1] * V0 - d[0] * V1) / (d[1] - d[0]);
        segment1 = (d[2] * V0 - d[0] * V2) / (d[2] - d[0]);
      } else if (d[1].val < 0.0) {
        segment0 = (d[0] * V1 - d[1] * V0) / (d[0] - d[1]);
        segment1 = (d[2] * V1 - d[1] * V2) / (d[2] - d[1]);
      } else  // d[2].val < 0.0
      {
        segment0 = (d[0] * V2 - d[2] * V0) / (d[0] - d[2]);
        segment1 = (d[1] * V2 - d[2] * V1) / (d[1] - d[2]);
      }
    } else if (negative == 2)  // and positive == 1
    {
      if (d[0].val > 0.0) {
        segment0 = (d[1] * V0 - d[0] * V1) / (d[1] - d[0]);
        segment1 = (d[2] * V0 - d[0] * V2) / (d[2] - d[0]);
      } else if (d[1].val > 0.0) {
        segment0 = (d[0] * V1 - d[1] * V0) / (d[0] - d[1]);
        segment1 = (d[2] * V1 - d[1] * V2) / (d[2] - d[1]);
      } else  // d[2].val > 0.0
      {
        segment0 = (d[0] * V2 - d[2] * V0) / (d[0] - d[2]);
        segment1 = (d[1] * V2 - d[2] * V1) / (d[1] - d[2]);
      }
    } else  // positive == 1, negative == 1, zero == 1
    {
      if (d[0].val == 0.0) {
        segment0 = V0;
        segment1 = (d[2] * V1 - d[1] * V2) / (d[2] - d[1]);
      } else if (d[1].val == 0.0) {
        segment0 = V1;
        segment1 = (d[0] * V2 - d[2] * V0) / (d[0] - d[2]);
      } else  // d[2].val == 0.0
      {
        segment0 = V2;
        segment1 = (d[1] * V0 - d[0] * V1) / (d[1] - d[0]);
      }
    }
    return true;
  }
  return false;
}
VectorXdual 3DTTIT(VectorXdual loc, bool& intersected) {
  VectorXdual res(2);
  Vector3dual V0 = loc.segment(0, 3);
  Vector3dual V1 = loc.segment(3, 3);
  Vector3dual V2 = loc.segment(6, 3);
  Vector3dual U0 = loc.segment(9, 3);
  Vector3dual U1 = loc.segment(12, 3);
  Vector3dual U2 = loc.segment(15, 3);
  Vector3dual S00;
  Vector3dual S01;
  Vector3dual S10;
  Vector3dual S11;
  if (intersects(V0, V1, V2, U0, U1, U2, S00, S01) && 
      intersects(U0, U1, U2, V0, V1, V2, S10, S11)) {
    Vector3dual uNormal;
    uNormal = (U1 - U0).cross(U2 - U0);
    Vector3dual vNormal;
    vNormal = (V1 - V0).cross(V2 - V0);
    Vector3dual D;
    D = uNormal.cross(vNormal);
    D.normalize();
    Vector3dual A;
    A = 0.25 * (S00 + S01 + S10 + S11);
    dual t00 = D.dot(S00 - A);
    dual t01 = D.dot(S01 - A);
    dual t10 = D.dot(S10 - A);
    dual t11 = D.dot(S11 - A);

    dual ta0 = min(t00, t01);
    dual ta1 = max(t00, t01);
    dual tb0 = min(t10, t11);
    dual tb1 = max(t10, t11);

    intersected = (ta1.val > tb0.val && ta0.val < tb1.val);

    res(0) = (ta0 - tb1) * 0.5;  // tc0
    res(1) = (ta1 - tb0) * 0.5;  // tc1
    return res;
  }
  intersected = false;
  return res;
}
\end{lstlisting}
\end{document}